\documentclass{webofc}
\usepackage[varg]{txfonts}   
\usepackage{natbib}

\begin{document}
\title{Streaming Readout of the CLAS12 Forward Tagger Using TriDAS and JANA2}
%
%

%
%

\author{
         \firstname{Fabrizio} \lastname{Ameli}\inst{5}
    \and
       \firstname{Marco} \lastname{Battaglieri}\inst{1,4}
        \and
        \firstname{Mariangela} \lastname{Bond\'i}\inst{4}
        \and    
        \firstname{Andrea} \lastname{Celentano}\inst{4}
        \and
        \firstname{Sergey} \lastname{Boyarinov}\inst{1}
        \and
        \firstname{Nathan} \lastname{Brei}\inst{1}
        \and
        \firstname{Tommaso} \lastname{Chiarusi}\inst{2}
        \and    
        \firstname{Raffaella} \lastname{De Vita}\inst{4}
        \and    
        \firstname{Cristiano} \lastname{Fanelli}\inst{6}
        \and
        \firstname{Vardan} \lastname{Gyurjyan}\inst{1}
        \and
        \firstname{David} \lastname{Lawrence}\inst{1}
            \fnsep\thanks{\email{davidl@jlab.org}}
        \and
        \firstname{Paolo} \lastname{Musico}\inst{3}
        \and
        \firstname{Carmelo} \lastname{Pellegrino}\inst{3}
        \and
        \firstname{Ben} \lastname{Raydo}\inst{1}
        \and
        \firstname{Simone} \lastname{Vallarino}\inst{4}
}

\institute{
    Thomas Jefferson National Accelerator Facility
\and
    Istituto Nazionale di Fisica Nucleare - Sezione di Bologna
\and
    Istituto Nazionale di Fisica Nucleare - CNAF
    \and
    Istituto Nazionale di Fisica Nucleare - Sezione di Genova
    \and
    Istituto Nazionale di Fisica Nucleare - Sezione di Roma
    \and
    Massachusetts Institute of Technology - M.I.T.  
}

\abstract{%
    An effort is underway to develop streaming readout data acquisition system for the CLAS12 detector in Jefferson Lab's experimental Hall-B. Successful beam tests were performed in the spring and summer of 2020 using a 10GeV electron beam from Jefferson Lab's CEBAF accelerator. The prototype system combined elements of the TriDAS and CODA data acquisition systems with the JANA2 analysis/reconstruction framework. This successfully merged components that included an FPGA stream source, a distributed hit processing system, and software plugins that allowed offline analysis written in C++ to be used for online event filtering. Details of the system design and performance are presented.
}
\maketitle
%
\section{Introduction}
\label{intro}
An effort was started in early Spring 2020 to develop a prototype streaming data acquisition system (DAQ) for the CLAS12 detector\cite{BURKERT2020163419} in experimental Hall-B at Jefferson Lab. This system brought together components from the existing CODA 
DAQ system\cite{Coda1}\cite{Coda2}, the TriDAS DAQ system\cite{Chiarusi_2017} and the JANA2 software framework\cite{Lawrence_2020}. The prototype system was used to successfully read out the CLAS12 Forward Tagger(FT) and Forward Hodoscope(FH) detectors in a streaming mode during an active beam test. The COVID-19 pandemic halted the beam operations early in the testing period, but the test was resumed once beam operations started back up in the summer. The longer term goals for the project are to expand the prototype to the full CLAS12 DAQ system and eventually deploy the system to other experiments at Jefferson Lab and elsewhere. One of the main benefits of a streaming readout (SRO) system is that it allows custom hardware triggering systems to be replaced with software algorithms run on cheaper commodity hardware. In addition to removing the deadtime inherent with traditional hardware triggers, this allows more complex triggering algorithms that can operate on whole detector events (as opposed to only fast subdetectors).

The following sections give details on the setup for the beam test and the various components of the prototype system. This is followed by some analysis results of the data taken with the system.

\section{Instrumentation and Resources}
\label{instrumentation}

\subsection{Experimental Setup}
The beam test focused on the reaction $e$ X $\to \pi^0$ $X$ with $\pi^0 \to 2 \gamma$  produced by the interaction of 10.6 GeV, $\sim$100 nA, CEBAF electron beam on 125$\mu$m lead and 40 cm gaseous deuterium targets. The inclusive $\pi^0$ electro-production was chosen because the two decay $\gamma$s  can be detected by a single detector (the CLAS12 Forward Tagger) reducing the complexity of the experimental set up. Moreover, the invariant mass of the two photons to form a $\pi^0$, provides a clean signature over the background due to the scattered electron and other electro-magnetic processes.

The Forward Tagger or FT\cite{Acker:2020brf} is part of the CLAS12 detector\cite{Burkert:2020akg} hosted in Hall-B at Jefferson Lab. It is composed of  a lead-tungstate electromagnetic calorimeter  (FT-Cal), used to measure the photon energy and position, and a plastic scintillator hodoscope (FT-Hodo), used to distinguish neutrals from charged  particles and, in turn, identify gammas. The FT covers a small solid angle ($0^o<\phi<360^o$ and $2.5^o<\theta<5^o$) in the beam direction. It is mainly used to detect electrons scattered at small angles from the target and forward-going neutral particles, such as $\pi^0$, produced in the interaction of the electron beam with the target. The 332  PbWO crystals of the FT-Cal and the  232 tiles of the FT-Hodo were read out by  JLab digitizers. The limited number of channels and the combination of two different detectors (calorimeter and plastic scintillators) usually used to trigger the experiment DAQ, represents the ideal bench test for a real  on-beam set up.  For a quantitative  assessment of the streaming read out DAQ chain, some data where also collected in standard triggered mode for later comparison.

\subsection{Readout Electronics}

In our streaming readout (SRO) system physics signals were continuously digitized by the fADC250 flash ADC. The fADC250 is a VME64x 16-channel direct-conversion ADC module, conforming to the VITA-41 switch serial standard (VXS). These high-speed flash ADCs were developed at JLAB as part of the 12 GeV upgrade. Currently these modules are deployed in many JLAB experiments, providing energy deposition, timing, as well as hit and trigger information. The fADC250 is equipped with an FPGA, that receives 12-bit data-words streaming at 250 MHz from 16 fADC channels in a module. The fADC250 FPGA performs data processing for each fADC channel, computes energy sum of all fADCs, and generates acceptance pulses for each fADC. Each VXS crate houses 16 fADC modules that are managed by the VXS Trigger Processor (VTP) module.
The VTP is a VXS switch card module, that was designed to play the leading role in the level1 trigger formation as a central or a global trigger processors (CTP, GTP) in a traditional triggered DAQ system.

\begin{figure}[htb]
\centering
\includegraphics[width=12cm,clip]{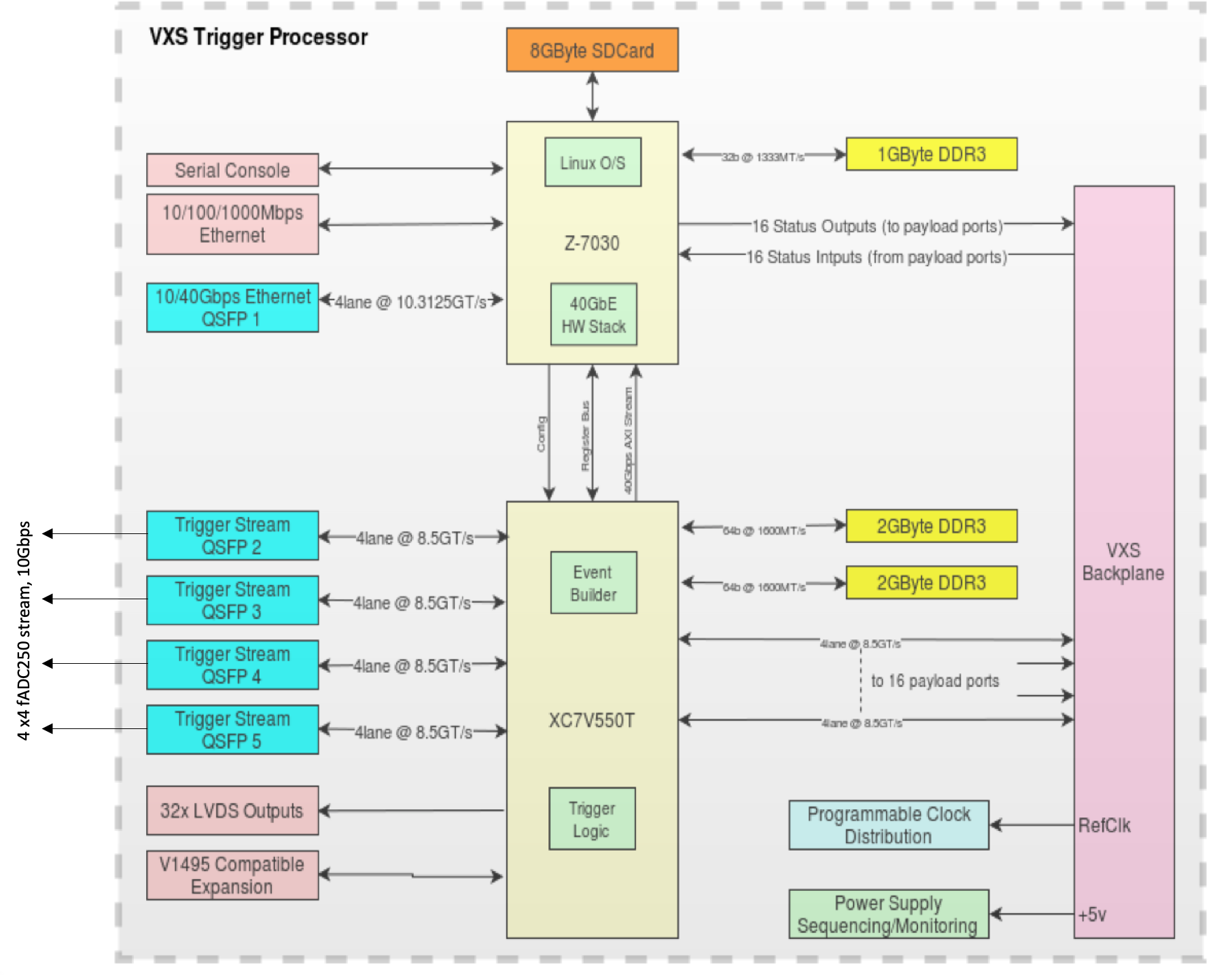}
\caption{VXS Trigger Processor (VTP) architecture.}
\label{fig:VTP architecture}
\end{figure}

Figure \ref{fig:VTP architecture} shows a diagram of the VTP module used to communicate and process data from the fADC250 modules in the crate. The design of the VTP contains high speed backplane serial links to each front-end payload module in the crate. Fiber optic serial links provide communication to other crates. In addition, the VTP has more FPGA resources for data processing logic, and a dual-core 1GHz ARM processor, capable of running data processing components, such as event building, trigger and processing diagnostics. These features make the VTP module an ideal candidate for designing a streaming data acquisition system. 

In the presented streaming DAQ system prototype the fADC250 data is received at 10Gbps from each payload slot of the VXS crate. Data from each fADC250 is then buffered into DDR3 memory, where each module has a dedicated 256MB space for data buffering. The role of these buffers is to allow a significant burst of physics input signals, as well as for handling substantial network delays or downstream processor latencies. The VTP streaming firmware implements 4 parallel instances of fADC250 streaming systems, each feeding a 10Gbps Ethernet link. Each instance handles 4 slots (i.e. 4 fADC250 modules), with a 1 GByte memory buffer. Every 8 slots of fADC share a 2GByte DDR3 buffer. Ethernet was chosen for the streaming readout interface because of its widespread support and compatibility. The VTP is programmed with the destination IP address and socket, where streaming data from 4 fADC slots are to be sent. Taking into account the computing power of contemporary servers, a socket/server per 10Gbps link is feasible for data transfer.
The fADC data-payload is packed into a TCP data frame, containing a header that includes information about the frame number and the timestamp. This information is necessary for ensuring the data coherency and synchronization. Streaming TCP frames correspond to a programmable time-span  (typically 65536ns) for which the reported fADC hits are collected. At the beginning of the streaming readout, the VTP module synchronously starts its own timer that is used to timestamp fADC hits. At every elapse of the frame-time, a TCP data frame is sent containing hits for the corresponding time-span.
VTP is responsible for dropping streaming data-frames in case the downstream receivers are not able to accept a higher input rate for a long periods of time. Burst conditions in the order of ~100msec at 32MHz/channel ( for all channels) are handled without data loss by the VTP DDR3 memory buffers. When the buffers are full, an entire frame (65536ns chunk of data) is dropped. These losses should not happen under normal conditions when the network and downstream processing chain are efficient enough to keep up with the readout rate. The VTP frame counter (the record-number in the TCP header) is used to identify the number of dropped frames, thus representing  the efficiency of entire data stream processing pipeline.

\section{Implementation}
\label{implementation}

\subsection{Front end Streaming Source}

The JLAB data acquisition system called CODA was designed to work with trigger-based readout systems. A key component is the Event Builder, which collects data from 100+ Readout Controllers (ROCs) and VXS Trigger Processor Boards (VTPs). In the traditional triggered mode of operation, the event builder builds events based on event number. Also used for triggered mode is the Trigger Supervisor (TS) module which synchronizes  all components using clock, sync, trigger and busy signals. ROCs would read front-end electronics over a VME bus, and VTPs are used to help form trigger decisions and report some trigger-related information. A detailed description of the CLAS12 triggered mode system can be found in the literature\cite{BOYARINOV2020163698}.

To use the available front-end electronics in streaming mode, the role of the TS was reduced to clock distribution, and the Event Builder was replaced with new SRO components and back-end software capable of gluing the front-end module information together based on timestamp instead of event number. In addition the role of the ROCS and VME bus were reduced to just the initial configuration of the front-end modules. In streaming mode, all front-end electronics readout is performed by the VTP boards over the VXS serial lines rather then VME bus. This allows an increase in the bandwidth limit from about 2GBit/s to 20Gb/s for each of the participating electronics crates, with the possibility to be increased to 40GBit/s if needed. New firmware was developed for the VTPs to implement streaming mode.

\begin{figure}[htb]
\centering
\includegraphics[width=8cm,clip]{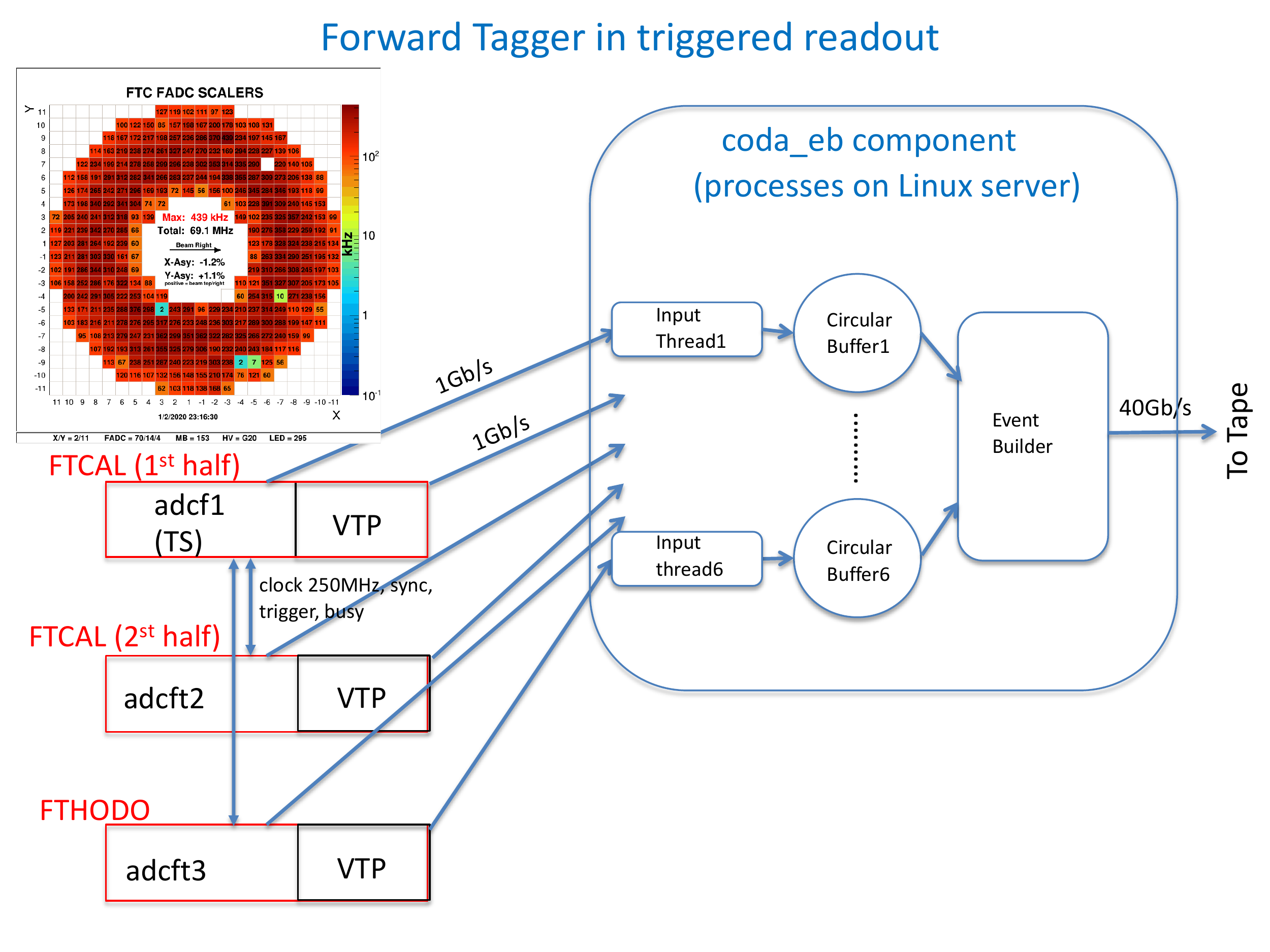}
\caption{Forward Tagger in triggered readout: Trigger Supervisor supplies 250MHz clock, sync and trigger signals to all front end components, and collects busy conditions}
\label{fig:triggered_mode}
\end{figure}

\begin{figure}[htb]
\centering
\includegraphics[width=8cm,clip]{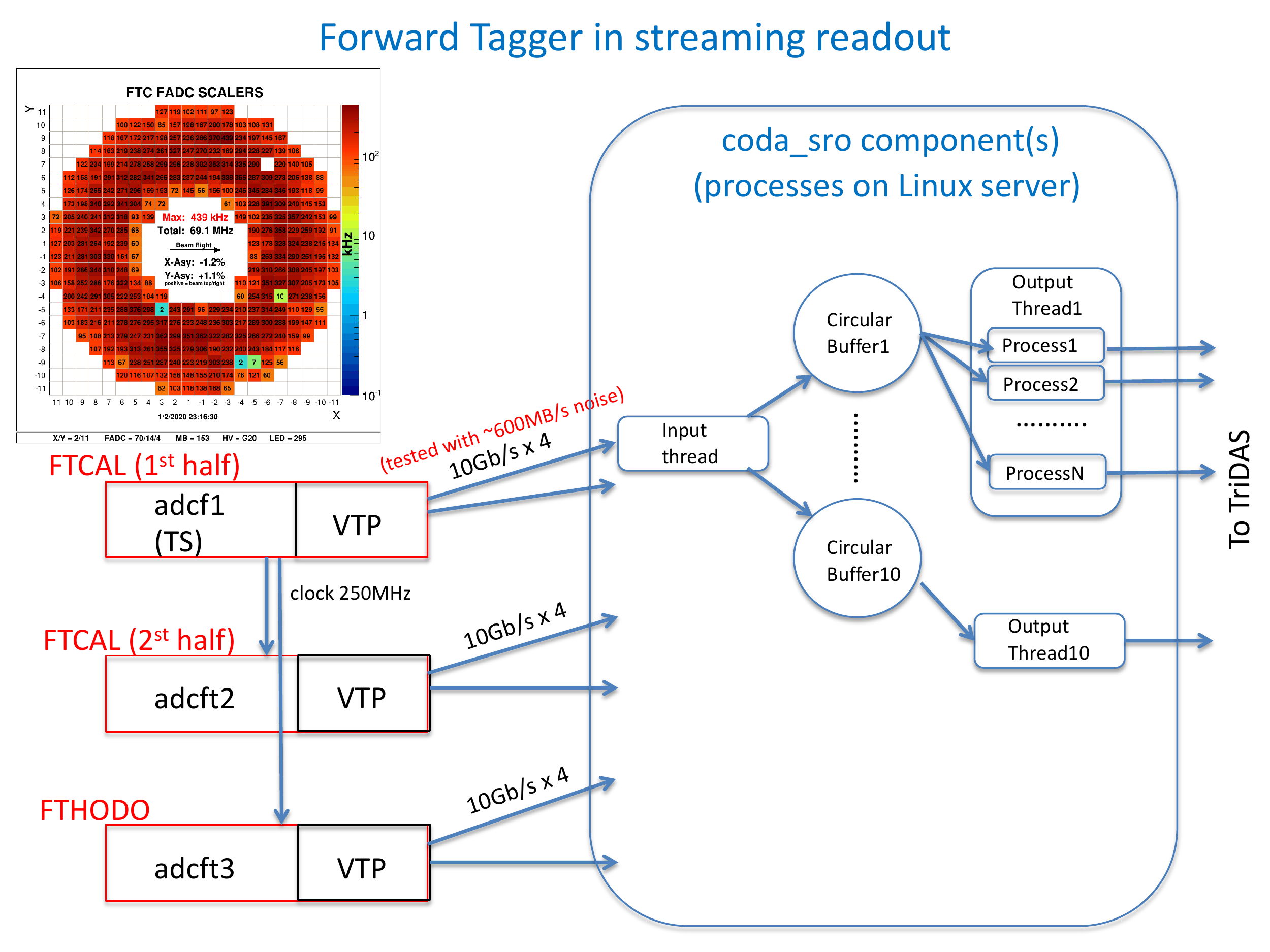}
\caption{Forward Tagger in streaming readout: Trigger Supervisor supplies 250MHz clock only, data is streaming freely from front end}
\label{fig:streaming_mode}
\end{figure}

Figures \ref{fig:triggered_mode} and \ref{fig:streaming_mode} show the original version of CODA, as well as the streaming version of CODA (without back-end which notes as TriDAS). In short, the front end readout software running inside the VME controllers and VTP boards was modified to stream data out freely, and a new SRO component was developed to be the intermediate translator between front-end and back-end. The SRO component is a multi-threaded and multi-node component capable of handling the 20Gb/s data rate from every electronics crate. It gets data from VTPs, converts it into TriDAS format applying appropriate format checks, and then supplies the results into the TriDAS.

\subsection{Online Streaming DAQ}
\begin{figure}[htb]
\centering
\includegraphics[width=8cm,clip]{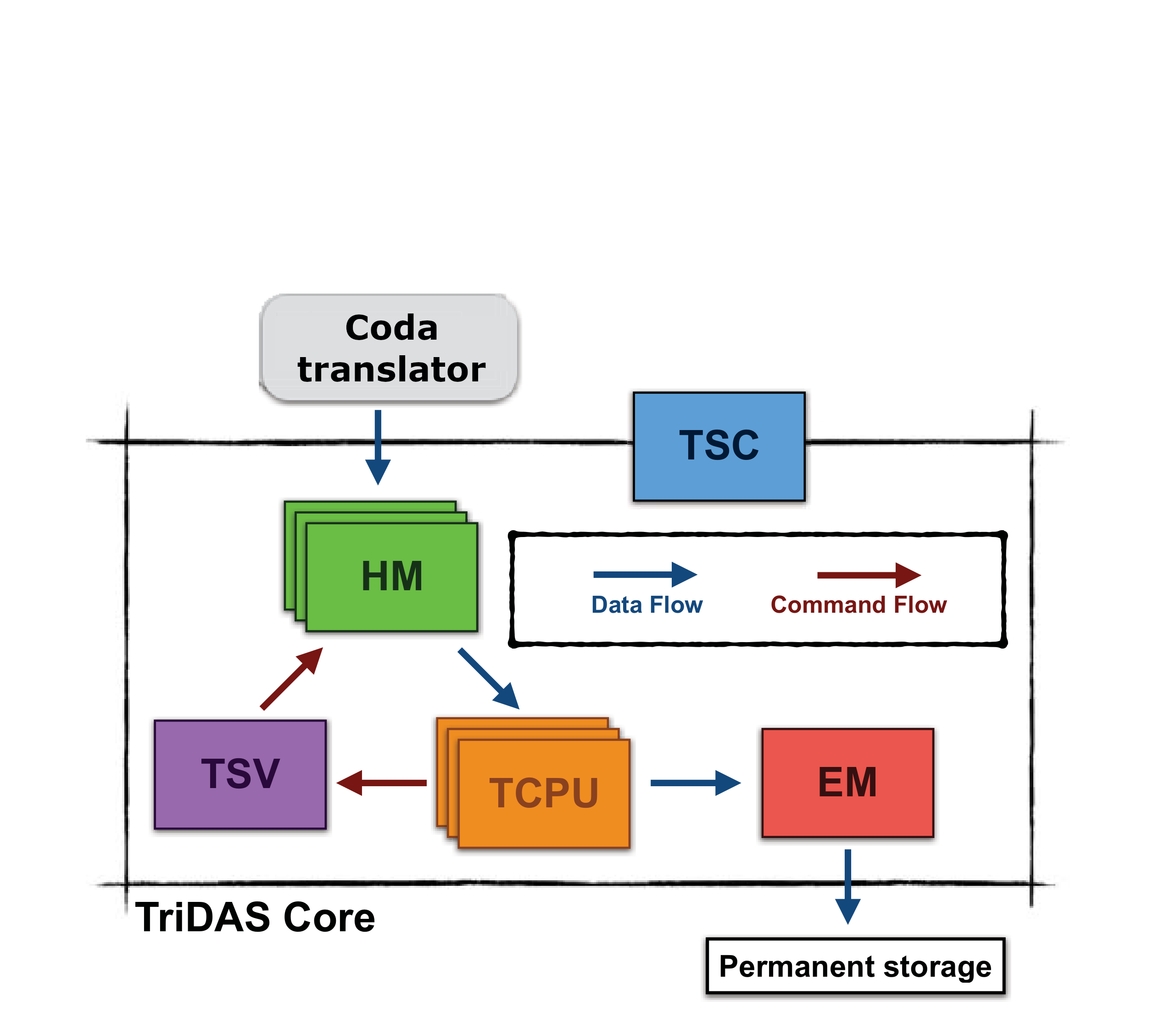}
\caption{The TriDAS flow diagram. Data flow from the translators (block on the upper-left corner) to the HitManagers (HM) where time-slicing takes place. Each time slice (TS) is coherently routed to a single TriggerCPU (TCPU) process for event building and classification/selection with JANA. The TriDAS SuperVisor (TSV) manages the load balancing via token passing. Selected events are stored to permanent media by the Event Manager (EM). The TriDAS Core application is steered by the TriDAS System Controller (TSC) which responds to the user commands throughout the TriDAS state machine.}
\label{fig:tridas}
\end{figure}

The Trigger and Data Acquisition System\cite{Chiarusi_2017} (TriDAS) is software originally designed and implemented for streaming read-out of Astroparticle Physics events, specifically for the NEMO project which aimed at developing the technologies to build a cubic-kilometre sized telescope for high-energy cosmic neutrinos. The TriDAS scalable, modular, and flexible design made it also adaptable to the requirements of a beam-based experiment with minimal development effort.

TriDAS is made by several software components, each devoted to a specific task in the data-processing chain  and implemented in C++11.

In this paper we describe the TriDAS as sketched in  fig. \ref{fig:tridas}, which represents the implementation realised for the CLAS12 streaming readout  test, in the summer of 2020. The HitManagers (HMs) represent the first data aggregation stage. They receive the data streams from a pre-defined number of CODA translators, topologically corresponding to a sector of the detector. Each HM subdivides the collected data into a sequence of time-ordered bunches of data, called Sector Time Slices (STS). The sequence of STSs reproduces the succession of time windows of fixed duration, typically 50\,ms. Sharing a common time reference, all HMs arrange their STSs according to the same intervals of time, which are referred to as Time Slices (TS).

The TriggerCPUs (TCPUs) receive the STS assembled by all HMs referring to the same TS and apply the event building and the classification/selection algorithms to the data (see section \ref{par:soft-trig}). Time Slices are processed in parallel by multiple threads of the same TCPU and by multiple TCPU processes running on the computing farm. For the CLAS12 FT and FH application, the Level 1 events (L1) consisted of the detector data within a time window of 200\,ns around a hit whose energy exceeded a threshold of $\approx$\,2\,GeV. No attempt is currently made to check for and recover events spanning two STS's due to the detector window (200\,ns) being so much smaller than the STS (50\,ms) that it represents a negligible amount of potential data loss.
Level 1 events identified within a TS are then fed to the L2 classification/selection algorithms that are implemented in separate binaries that are specified in the run configuration file. These binaries are loaded and configured at run-time, allowing one to easily change the L2 algorithms or their parameters on a run-by-run basis without the need for recompiling while still keeping the highest possible computation efficiency.

A token-based mechanism is at the base of the TriDAS SuperVisor (TSV) load balancing. Each TCPU thread owns a token that is given to the TSV on completion of the TS processing. The TSV, then, maintains a pool of ``free to use'' TCPU threads which are then matched to the new Time Slices that are continuously assembled by the HMs.

The Event Manager (EM) collects the selected L2 events and then writes them to the so called Post Trigger (PT) file.

\begin{figure}[htb]
\centering
\includegraphics[width=8cm,clip]{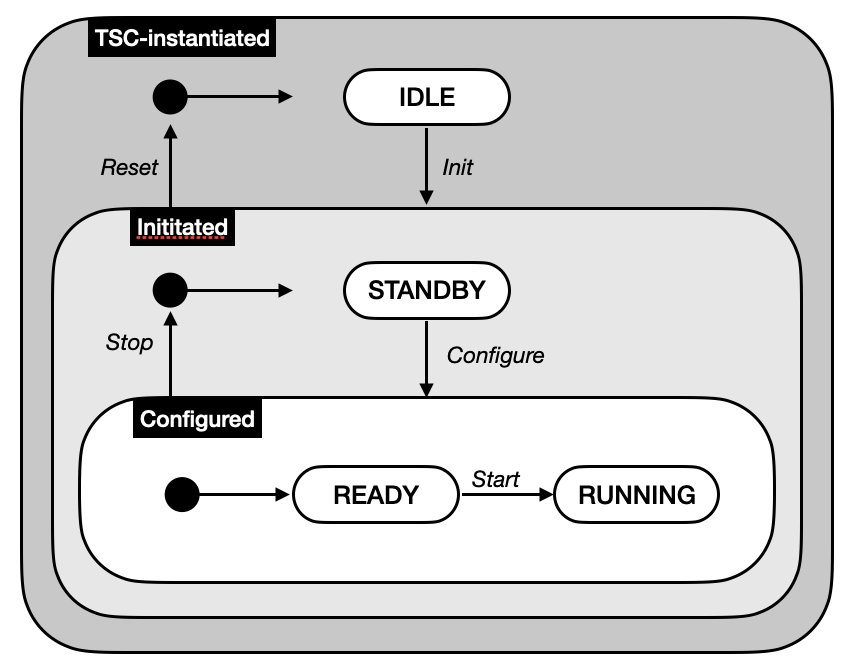}
\caption{The TriDAS state machine. See text for details.}
\label{fig:tridas_state_machine}
\end{figure}

The TriDAS System Controller (TSC) is the part of the system with which users directly interact. Through it, users may configure and control the TriDAS activities. For the aforementioned test with CLAS12, a simple interface to the TSC was built in order to steer TriDAS along the hierarchical state machine sketched in fig. \ref{fig:tridas_state_machine}. 

In the IDLE state, only the TSC process is running and waits for user commands. Upon the \textit{Init} transition, the TSC retrieves a JSON-formatted run configuration file, called \textit{Datacard}. The Datacard describes the geometry of the detector and the configuration of the TriDAS system for a given run. If the transition is successful, the state machine moves into the Initiated sub-state machine. TriDAS is now in the STANDBY state, where still no other process than TSC is running. During the  \textit{Configure} transition, The TSC decides the run number and then starts the HM, TCPU and EM processes on the corresponding nodes. If all the processes start successfully the state machine moves into the Configured sub-state machine. The TriDAS is now in the READY state. During the \textit{Start} transition, the TSC computes the start date and time of the run (which for the CLAS12 case is always the fixed value 01/01/2020 00:00:00) and starts the TSV.  If this transition is  successful, TriDAS moves into the RUNNING state.

\subsection{Software Trigger}\label{par:soft-trig}

The TriDAS system supports user-level plugins to allow implementation of custom processing algorithms which can be used to implement a software trigger. For this prototype system, a TriDAS plugin was constructed that implemented the JANA2 framework. JANA2 is a multi-threaded event processing/analysis framework designed for both offline and streaming applications. User algorithms written within the JANA2 framework were then made available for forming software triggers in the form of JANA2 plugins. The benefit of this is that the full suite of reconstruction algorithms used in the offline reconstruction are available for use as triggers/filters in the streaming system. This includes accessing translation tables, and calibration constants.

The software triggering itself was done by using multiple JANA plugins, each implementing their own trigger(s). Each plugin produced one or more \textit{TriggerDecision} objects for each potential ``event'' identified by the TriDAS system. The decision for each algorithm was in the form of a 16bit integer where a value of zero meant \textit{no-keep} and any non-zero value meant \textit{keep}. If any trigger algorithm indicated a \textit{keep} condition then The TriDAS system was told to keep the event. A unique 16bit ID was assigned to each trigger algorithm (passed in the \textit{TriggerDecision} object). The 16bit ID and 16bit decision for each ``event'' was given to TriDAS so it could store the decision for each algorithm with each event written out. 

\begin{figure}[htb]
\centering
\includegraphics[width=11cm,clip]{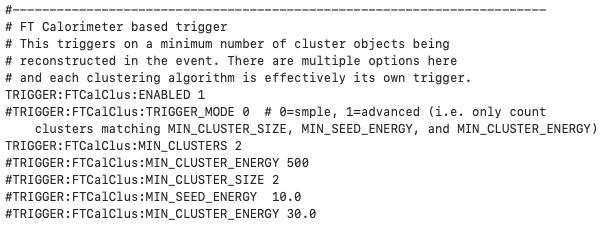}
\caption{Snippet from the JANA configuration file showing settings for one of the software triggers implemented. The format is simple key-value pairs. Lines starting with ``\#'' are comments. Having the keys start with ``TRIGGER:FtCalClus:'' was just a choice of convention for this particular beam test.}
\label{fig:JANA_config_snippet}
\end{figure}

The JANA2 plugins list was determined by the JANA configuration file. Configuration settings for the individual triggers were also set in this file. The system allows different individuals to maintain the code base for their trigger separately while the selection of which triggers are used and what their configurations are is kept in a single configuration file that is read in at run time. Figure \ref{fig:JANA_config_snippet} shows a snippet of the configuration file with settings for the FT calorimeter multi-cluster trigger.

\begin{figure}[htb]
\centering
\sidecaption
\includegraphics[width=9cm,clip]{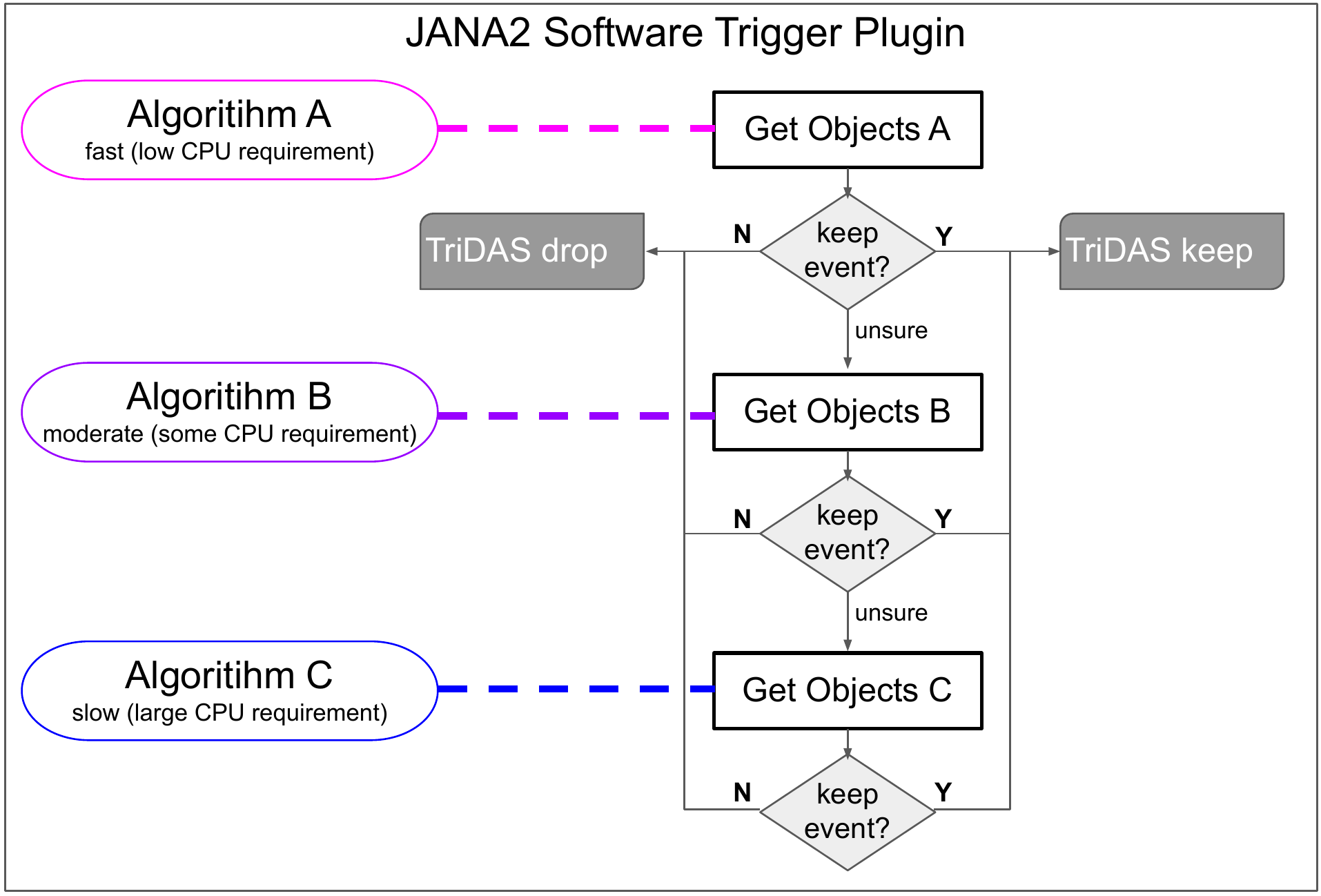}
\caption{Illustration of how the JANA2 on-demand design can be leveraged to reduce the overall CPU required to implement a software trigger. In this scenario, the more expensive algorithms are only run on an event/time slice when a keep or drop decision cannot be made using a less expensive algorithm.}
\label{fig:JANA2_example_trigger_diagram}
\end{figure}

The on-demand design of JANA2 specifically supports multi-tiered triggering. This means trigger algorithms can be designed such that more expensive algorithms are only run for events or time slices when a decision cannot be made using the output of less expensive algorithms. The benefit of this is that the compute resource required for the software trigger can be provisioned for the average time needed for a keep/no-keep decision rather than for the most expensive algorithm. For example, consider a situation in which one wishes to trigger on events with a detected proton track and two other charged tracks in the forward direction such as a rare Primakoff reaction $\gamma p\rightarrow p\pi^{+}\pi^{-}$. Even the rough tracking algorithm used to identify the two very forward going tracks can be expensive. At the same time, one wants to take a significant amount of of pre-scaled events using a minimum bias trigger where only a hit count or minimal calorimeter energy is needed. This would be a very fast algorithm. JANA2 can be configured to only run the more expensive Primakoff tracking trigger for those events which were not already flagged for saving by the fast minimum bias trigger. Figure \ref{fig:JANA2_example_trigger_diagram} illustrates how this can work. 

\section{Data Analysis}
\label{analysis}

\begin{figure}[htb]
\centering
\sidecaption
\includegraphics[width=9cm,clip]{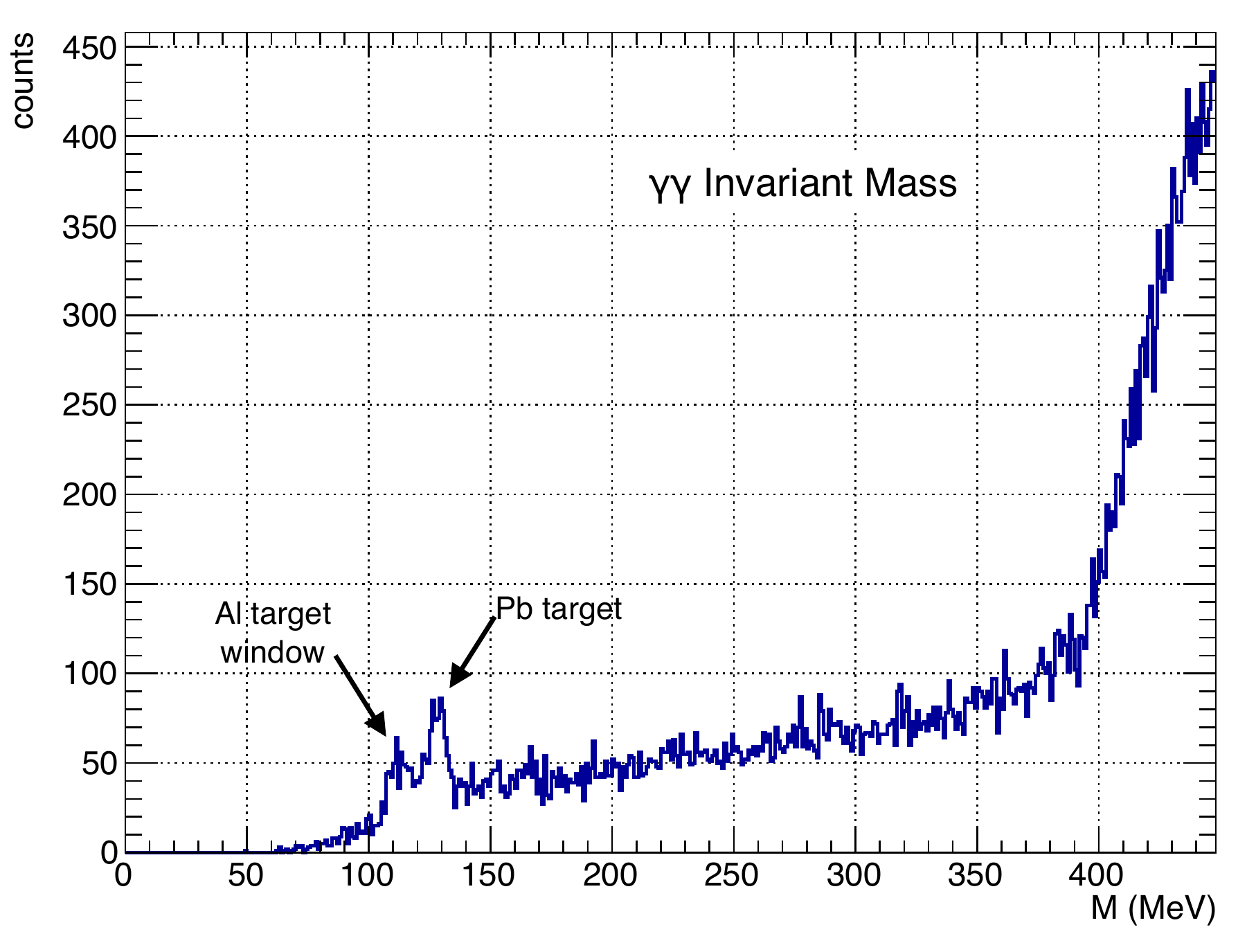}
\caption{$\gamma$$\gamma$ invariant mass spectrum. The labeled peaks are both due $\pi^\circ\rightarrow\gamma\gamma$ decays. The peak marked \emph{Al target window} has its position shifted to a lower invariant mass due to the assumption that the vertex is located at the Pb target position when calculating the invariant mass.}
\label{fig:gamma_gamma_invariantMAss}
\end{figure}

The off-line data analysis is focused on identification of the $\pi^{\circ}$->$\gamma\gamma$ decay events where both photons are detected in the calorimeter. In particular, in this paper we report the result obtained by the electron-beam on lead target test.\\
In the off-line data reconstruction, performed by applying the same full suite of reconstruction algorithms used in the on-line reconstruction, the recorded signal of each crystal was converted into energy by applying proper calibration constants. The latter were determined in a previous calibration run performed in standard trigger mode. The standard calibration procedure is described in \cite{ACKER2020163475}.\\
Figure \ref{fig:gamma_gamma_invariantMAss} shows the reconstructed $\gamma$$\gamma$-invariant mass spectrum. It is characterized by two peaks on the $\pi^{\circ}$ mass region: the first peak at higher mass is associated to $\pi^{\circ}$ production from the lead target, while the second one is related to $\pi^{\circ}$s production from the aluminum target window.

\subsection{Artificial Intelligence}
\label{sub:ai}

SRO can further the convergence of online and offline analyses allowing to incorporate new emerging software approaches. For example, the inclusion of high level A.I. algorithms in the analysis pipeline can foster better data quality control during data taking and shorter analysis cycles.
A.I.  is  becoming  ubiquitous  in  nuclear  and particle physics and  encompasses  all  the  concepts  related  to  the  integration of intelligence into machines; unsupervised learning is a type of algorithm able to learn patterns from untagged data (\textit{i.e.}, no training phase) offering new solutions to near real-time reconstruction problems.

An unsupervised hierarchical clustering algorithm inspired by hdbscan \cite{campello2015hierarchical} has been developed as a plugin within the JANA2 framework.  
The following are essential features of this unsupervised approach: (i) can be easily ported to other experiments; (ii) formally does not depend on cuts making it less sensitive to variations in experimental conditions during data taking; (iii) able to cope with large number of hits; (iv) excels when dealing with challenging topologies/arbitrary shaped clusters, different cluster size and noise. 

The main idea behind the hierarchical clustering is to consider all the information at the hit-level in the detector (spatial, time, and energy) and look at the density of the hits in that space of parameters, after defining a metric (\textit{e.g.}, euclidean) which allows to define what is called ``mutual reachability'' among points. In this way, clusters can be interpreted as more likely (higher density) regions separated by less likely regions (lower density). Within this framework all hits have a probability of belonging to a cluster as well as of being outliers and one can make decisions when forming clusters based on these probability values.

Tests have been performed both online and offline (on collected data) to analyze and reconstruct clusters in the FT-Cal, and provided results consistent with the $\pi^{\circ}$ yields already discussed.

\section{Summary}
\label{results}

A prototype streaming data acquisition system was successfully tested in beam conditions during the summer of 2020. The prototype system combined several different software systems with an existing hardware DAQ system for the test. These included the CLAS12 detector and CODA DAQ system, the TriDAS streaming DAQ system, and the JANA2 event processing framework. The test successfully read out the CLAS12 Forward Tagger detector and subsequent analysis was able to extract a clean physics signal in the form of a $\pi^\circ$ invariant mass peak. The prototype system is being used as the basis for developing a larger system planned for the entire CLAS12 detector and its future physics program.

\section{Acknowledgements}

We would like to acknowledge the CLAS12 collaboration as well as the JLab technical staff for their accommodation and support of this effort. 

The INFN Group has been supported by Italian Ministry of Foreign Affairs (MAECI) as Projects of
great Relevance within Italy/US Scientific and Technological Cooperation under grant n.
MAE0065689 - PGR00799.

This material is based upon work supported by the U.S. Department of Energy, Office of Science, Office of Nuclear Physics under contract DE-AC05-06OR23177.
\bibliography{references}

\end{document}